\documentstyle[aps,twocolumn]{revtex}
\input epsf.sty
\begin{document}
\twocolumn[\hsize\textwidth\columnwidth\hsize\csname@twocolumnfalse\endcsname

\draft
\title {Avoided Critical Behavior in O(n) Systems}
 \author {Zohar Nussinov, Joseph Rudnick, Steven A.
  Kivelson} \address{Department Of Physics, University Of California,
  Los Angeles, CA 90024} \author{L.N.Chayes} \address{Department Of
  Mathematics, University Of California, Los Angeles, CA 90024}
\date{\today} \maketitle
\begin{abstract}
  Long-range frustrating  interactions, even if their
  strength is infinitesimal, can give rise to a dramatic proliferation of
  ground or near-ground states. As a consequence, \mbox{the ordering
    temperature} can exhibit a 
    discontinuous
  drop as a function of the frustration.  
  A simple model of a doped Mott insulator, where the 
short-range tendency of holes to phase separate competes with 
long-range Coulomb effects, exhibits this 
``avoided critical''
behavior. This model 
may serve as a paradigm for many other systems.    
\end{abstract}

]



A wide variety of systems display equilibrium domain patterns
characterized by periodic (or nearly periodic) variations of an 
order parameter.  These patterns are stabilized by competing interactions.
Linear arrays of stripes and hexagonal arrays of 
bubbles are ubiquitous in thin films of magnetic garnets, and 
ferrofluids. Similar morphologies are 
also seen in Langmuir films, membranes, semiconductor surfaces, and 
many
other systems in which an otherwise uniform ground state is thwarted 
by a competing ``frustrating'' interaction of one sort or another$^{[1]}$.
Lately, stripe structures have been detected $^{[2]}$ in doped 
Mott insulators, including the high $T_{c}$ superconductors: the 
ordered states in these compounds consist of arrays of charged 
stripes which form antiphase domain walls between 
antiferromagnetically ordered spin domains. In the absence of a 
frustrating  
Coulomb interaction ({\it i.e.} for neutral holes), a lightly doped
Mott insulator is unstable to phase separation into a hole-rich 
``metallic'' phase and a hole-deficient antiferromagnetic phase.
Electrostatic repulsions forbid macroscopic charge seperation;
the compromise leads to the formation of stripe morphologies
on an intermediate scale. 
 
In this letter, we shall explore the effect of fluctuations on the
periodic 
structures in a simple model  of uniformly frustrated $O(n)$ spins.
We study the problem using two complementary approaches:
a low temperature, spin-wave 
expansion, and a perturbative expansion for large $n$, which
we carry through to order $1/n^{2}$.
It is found that the ordering 
temperature, $T_{c}(Q)$, as a function of the strength of the
frustrating interaction ``$Q$'' may, in 
certain instances, satisfy the inequality
\begin{eqnarray}
T_{c}(Q=0) > \lim_{Q \rightarrow 0} T_{c}(Q).
\end{eqnarray}
In other words, an infinitesimal amount of frustration 
depresses the ordering temperature discontinuously!
Specifically, we shall argue on the basis of a low 
temperature expansion about stripe like ground states,
that, in the absence of lattice effects, $T_{c}(Q)=0$ for $n > 2$ 
and $Q>0$.  Lattice anisotropies elevate $T_{c}(Q)$ from zero, but the 
discontinuity in $T_{c}(Q)$ persists for $2< d \le 3$ and $n > 2$.
For $n=2$, the lower critical dimension is three, and here
the finite temperature 
``ordered'' phase 
exhibits power law decay of correlations;
the model with $n=2$ has the same hydrodynamic description as a 
smectic liquid crystal.

For low frustration $Q$, the behavior of the system is controlled
by the proximity to the ``avoided critical temperature'' 
$T_{c}(Q=0)$.  
The following picture emerges of the thermal evolution of the model, 
as summarized in Fig. 1: At temperatures somewhat 
above
$T_{c}(Q=0)$, {\it{two large lengths}} govern the exponential decay
of correlations.
As the temperature is lowered below a crossover temperature
$T_{1}(Q) \sim T_{c}(Q=0)$, the system enters a low temperature regime
characterized by an oscillatory spin structure function
with a 
{\it{single length}} controlling the exponential decay of 
correlations at long distances. 
$T_{1}(Q)$ is a ``disorder line'' in the sense that
as $T$ approaches  $T_{1}$ from below, the wave-length of the oscillations
diverges, but no 
phase transition occurs.
As $T$ is lowered further, the wave-length decreases until,
as $T \rightarrow T_{c}(Q)$, it smoothly
approaches the period of the ordered phase that appears below 
$T_{c}(Q)$. However, an additional crossover occurs at a temperature
$T_{2}(Q)$ which lies between $T_{c}(Q)$ and $T_{1}(Q)$, such that
for $T_{2}(Q)\gg T\gg T_{c}(Q)$, there are again two
long lengths characterizing the fall-off of correlations, where the
new length is akin to the Josepshon length in the ordered phase of 
the unfrustrated system.  This second length can only be seen in
the context of a $1/n$ expansion.
The existence of multiple correlation and modulation
lengths is a common feature of the physics of all the various
frustrated systems alluded to above $^{[1]}$. A finite size scaling analysis,
which is a simple extention of an argument presented previously$^{[9]}$
allows us to identify the longest length in the temperature regime above
$T_{1}$ and below $T_{2}$ as a ``domain'' size, $R$, within which 
the physics is essentially that of the unfrustrated system, and to extract the
scaling relation (which is reproduced by the large $n$ results)
\begin{eqnarray}
R \sim \sqrt{Q/\xi_{0}}
\end{eqnarray}
where $\xi_{0} \sim [T_{c}(Q=0)-T]^{-\nu}$ is the correlation length
in the unfrustrated system at temperature $T$.


\noindent{ \bf  The Coulomb Frustrated Ferromagnet:}
As a concrete example, we consider a system with a short-range 
tendency to phase separation which is frustrated by a long-range 
Coulomb interaction.
A simple spin Hamiltonian which represents these 
competing interactions is 
\begin{eqnarray}
H_{0} = - \sum_{<{\vec  x},{\vec  y}>} S({\vec  x}) S({\vec  y}) +  
\frac{Q}{2} \sum_{{\vec  x} \neq {\vec  y}} \frac{S({\vec  x}) 
S({\vec  y})}{|{\vec  x}-{\vec  y}|}.
\end{eqnarray}
Here, $S({\vec  x})$ is  a coarse grained scalar variable which represents the 
local charge density. Each site ${\vec  x}$ lies on a 
cubic lattice (of size $N$) and represents a small region of space in 
which
$S({\vec  x})>0$, and $S({\vec  x})<0$ correspond to the positively and 
netatively charged
phases respectively.  
The first ``ferromagnetic'' term represents the 
short-range
(nearest-neighbour) tendency to phase-separation, 
while the second term is the 
Coulomb interaction.
Non-linear terms in the full Hamiltonian typically fix 
the locally preferred values of $S({\vec  x})$. 
One may consider $d \neq 3 $ dimensional variants
wherein the spins lie on a hypercubic lattice, and the Coulomb kernel
in $H_{0}$ is replaced by $Q |{\vec  x}-{\vec  y}|^{2-d}$.
$H_{0}$ can be fourier transformed as
\begin{equation}
H_{0}=\sum_{\vec k} J(\vec k) |S({\vec k})|^{2}
\label{eq:h0ofk}
\end{equation}
where the kernal
\begin{equation}
J(\vec k) =\frac{1}{2}[ Qk^{-2} + r_{0} + k^{2} +  \ldots ]
\label{eq:jofk}
\end{equation}
where $ r_{0}=-2d$, and  the ellipsis represents higher order terms in powers 
of $k$.  We will neglect these terms for now, as they are unimportant 
in the contiuum;  however, we will need to include some of these
terms when we treat lattice effects since they are the ones that reduce
the full rotational symmetry of free space to the point group 
symmetry of the lattice.\cite{1}
\begin{figure}[htb]
\centerline{\epsfxsize=2.8in \epsffile{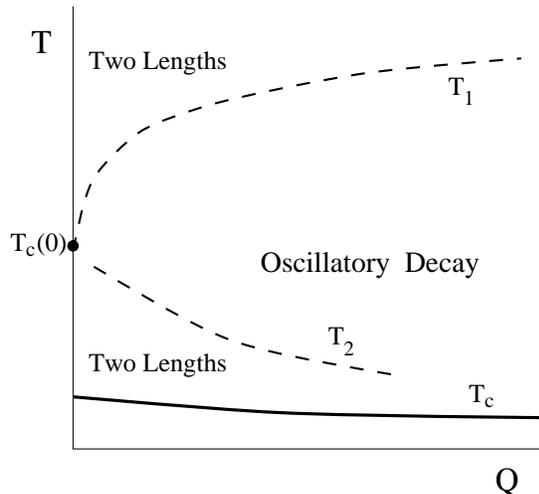}}
\caption{Schematic Phase Diagram. $Q$ is the strength of the frustration. 
For $T>T_{1}(Q)$ and $T_{2}(Q) > T > T_{c}(Q)$, 
there are two long lengths governing the fall of correlations.
For $T<T_{1}(Q)$, the spin-spin correlation funcition exhibits long distance
oscillatory structure. 
The thick black dot marks $T_{c}(Q=0)$, the 
ordering temperature in the absence of frustration;
this is what we term ``the avoided critical point''.
In the continuum limit, $T_{c}(Q>0)=0$; lattice effects 
result in a non-zero, yet small $T_{c}(Q)$.}
\label{fig1}
\end{figure}

We now generalize  this model, allowing the spins to
have $n$ components, and replacing all two spin 
products in $H_{0}$ with a scalar product.
We treat both the ``soft-spin'' version of this model, in which
we include the non-linear interaction
\begin{eqnarray}
H_{soft} = H_{0} +  u \sum_{{\vec  x}} [{\bf   S}^{2}({\vec  x})-1]^{2}
\end{eqnarray}
with $u>0$,
or the ``hard-spin'' version, which can be viewed as the 
$u\rightarrow \infty$ limit of the soft-spin model, in which we 
instead enforce the local constraint, $|{\bf   S}({\vec  x})|=1$.

When $n \ge 2$, we can construct a set 
of ground-state configurations which are simple spirals of the form
\begin{eqnarray}
{\bf   S}^{g}({\vec  x}) ={\bf   a} \cos(\vec{k}_{\min} \cdot \vec{x}) + 
{\bf   b} \sin( \vec{k}_{\min} \cdot \vec{x}),
\end{eqnarray}
where ${\bf  k}_{\min}$ denotes any which wave-vector which minimizes 
the  interaction kernel, $J(\vec k)$, in $H_{0}$, and the prefactors
satisfy
\begin{eqnarray}
{\bf   a}\cdot {\bf   b}=0 ; ~~{\bf   a} \cdot {\bf   a} = 
{\bf   b}\cdot {\bf   b} =1.
\end{eqnarray} 
It is readily seen that such states are unstable to transverse 
fluctations. One may expand $H_{soft}$ to quadratic order in 
fluctuations about the ground state $\Delta {\bf S}({\vec  x}) = 
[{\bf S} ({\vec  x}) - {\bf S}^{g}({\vec  x})]$, 
and estimate the thermal average of $\Delta {\bf  S}^{2}({\vec  x})$. 
For $d=3$ and
a vanishing lower cutoff $\epsilon$ on $||{\bf  k}|-|{\bf  k}_{\min}||$:
\begin{eqnarray}
\frac{<(\Delta \bf  {S})^{2}>}{T} = \frac{(n-2) \sqrt{Q}}{4 \pi^{2} 
\epsilon} - \frac{1}{16 \pi} Q^{1/4} \ln |\epsilon|.
\end{eqnarray}
For $n>2$, the leading order divergence is $O(\epsilon^{-1})$, 
indepepent of $d$; 
for $XY$ spins ($n=2$) the leading order divergence, in $2 \le d < 3$,
is $O(\epsilon^{d-3})$. 
In $d=3$, $XY$ spins exhibit power law correlations at low temperatures.
As we shall show, for small $Q$ and
$n=2,3$, these simple spirals are the only possible
ground states;
for $n\ge 4$  other types of ``multi-spiral''
ground-states are possible.  


\noindent{\bf The Three-Dimensional Spherical Model:}
To make the phase diagram non-trivial, yet tractable,
we may solve the scalar spin model subject to the single mean 
spherical constraint$^{[1,5]}$ 
\begin{eqnarray}
\sum_{{\vec  x}} <S^{2}({\vec  x})> = N.
\label{eq:spherical} 
\end{eqnarray}
It is known  ${[11]}$ that, in many respects, 
the spherical model is equivalent to the $ n \rightarrow \infty$ limit of the 
$O(n)$ model. Here, the effective Hamiltonian is the same $H_{0}$
defined above, with $r_{0}\rightarrow r$, where
$r$ is a Lagrange multiplier determined implicitly from
the constraint equation (\ref{eq:spherical}).
In order that all modes have a bounded Boltzmann weight
it is necessary that $r \geq -2\sqrt{Q}$.
By equipartition, $<|S(\vec{k})|^{2}>
= T[k^{2}+Q/k^{2}+r]^{-1}$, so the mean spherical constraint reads
\begin{eqnarray}
\frac{1}{T} = \int_{|{\vec{k}|<\Lambda}} \frac{d^{3}k}{(2 \pi)^{3}} 
~\frac{1}{k^{2}+Qk^{-2}+r},
\label{eq:roft}
\end{eqnarray}
where $\Lambda$ is an ultraviolet cutoff.
If this equation cannot be satisfied for any value of $r \geq -2\sqrt{Q}$,
then the system is at or below criticality.

Observe, from Eq.(\ref{eq:roft}), that for $Q > 0$, the integral 
diverges when  $r \rightarrow -2 \sqrt{Q}$.  Thus, the constraint 
can be satisfied for any non-zero $T$:
$T_{c}(Q>0)=0$.
By contrast, when $Q=0$ ( which is
the standard three-dimensional short-range ferromagnet) ~$T_{c}$ is non-zero. 
Thus a discontinuity in $T_{c}(Q)$ 
is seen to exist.
However, even though $T_{c}=0$ for $Q > 0$, there is a genuine
zero temperature phase transition with the usual
 $n\rightarrow \infty$ critical 
exponents, {\it e.g.}
$\nu=1$ and $\gamma=2$ in $d=3$.  As we shall see,  lattice
effects elevate $T_c(Q)$, but do not change the critical properties,
nor, in $d=3$, eliminate the discontinuity.

The pair correlator is given by 
\begin{eqnarray}
G({\vec  x}) = && \frac{1}{(2 \pi)^{3}} \int d^{3}k ~<|S({\vec k})|^{2}>~ 
\exp[i {\vec k} \cdot {\vec x}] \nonumber 
\\ = && \frac{T}{2 \pi^{2}|{\vec  x}|}\int_{0}^{\infty} dk
  \frac{k^{3}
    [Im\{{e^{ik|{\vec  x}|}\}]}}{(k^{2}+\alpha^{2})(k^{2}+\beta^{2})}
\end{eqnarray}
where
\begin{equation}
  \alpha^{2}, \beta^{2} = \frac{r \mp \sqrt{r^{2}-4Q}}{2}.
\end{equation}
When $r > 2 \sqrt{Q}$, the integral can be readily evaluated
by applying the residue theorem to the poles lying on the imaginary axis at $k=\pm \imath \alpha ,\pm \imath \beta$, 
\begin{eqnarray}
  G({\vec  x}) = && \frac{T ( \beta^{2} e^{-\beta |{\vec  x}|} 
  - \alpha^{2} e^{-\alpha |{\vec  x}|} ) }{4 \pi |{\vec  x}| 
  (\beta^{2}-\alpha^{2})}. 
\label{eq:g}
\end{eqnarray}
Note the existence of two macroscopic correlation lengths -- 
a consequence of charge neutrality: In $H_{0}$, the spins portray charges, and therefore 
must sum to zero,
\begin{equation}
\int G({\vec  x})~d^{3}x=< |\int S({\vec  x})~d^{3}x|^{2}> =0.
\end{equation}
Whenever $G$ is dominated by its long-distance behavior, the integral 
can vanish only if $G(x)$ contains positive and negative 
contributions, as in Eq.(\ref{eq:g}).
The latter integral can be made to vanish only if $G({\vec x})$ 
contains, at least, two length scales. 
At high temperatures, the length
\begin{eqnarray}
\xi_{1} \equiv |Re \{\beta\} |^{-1} \approx  r^{-1/2} (\mbox{ 
for $r \gg 2 \sqrt{Q}$}). 
\end{eqnarray}
plays the role of the correlation length of the canonical short-range 
ferromagnet ({\it i.e.} 
with $Q=0$).  
Note that now, however, an additional correlation length appears: 
\begin{eqnarray}
\xi_{2} \equiv |Re \{\alpha\} |^{-1} \approx {Q^{-1/2}}/{\xi_{1}} .
\end{eqnarray}
Thus, $\xi_2 \gg \xi_1$ in the limit of weak frustration, $Q \ll 1$.    
The analytic continuation of Eq.(\ref{eq:g}) to low temperatures, $r< 2 \sqrt{Q}$, is
\begin{eqnarray}
 G({\vec  x}) = && \frac{T}{4 \pi} \exp({-\alpha_{1}|{\vec  x}|}) \nonumber 
\\
 && \times \left[\frac{(\alpha_{2}^{2}-\alpha_{1}^{2})\sin \alpha_{2} 
|{\vec  x}|
      + 2 \alpha_{1} \alpha_{2} \cos
      \alpha_{2}|{\vec  x}|}{4\alpha_{1}\alpha_{2} |{\vec  x}|}\right]
\end{eqnarray}
where $\alpha \equiv \alpha_{1}+ \imath \alpha_{2}$. 
The temperature $T_{1}$, defined by $r(T=T_{1})=~2 \sqrt{Q}$, 
marks a dramatic crossover. 
At low temperatures ($ T < T_{1}$), the system possesses a
single correlation length $\xi = |\alpha_{1}|^{-1} =
2 {[r+ 2 \sqrt{Q}]}^{-1/2}$, and a single modulation length 
$L_{D} = 2 \pi / |\alpha_{2}| = 4 \pi {[-r+ 2 \sqrt{Q}]}^{-1/2}$; at high temperatures 
($T>T_{1}$), the system possesses two distinct correlation lengths.  
When $T=T_{1}^{-}$, the modulation length diverges as 
$L_{D} \sim (T_{1}-T)^{-1/2}$.
Many quantities of interest  ({\it e.g.} the specific heat 
$C_{V}$)
albeit analytic, display a
crossover at $T_{1}(Q)$. As $Q$ tends to zero,  the crossover 
temperature $T_{1}(Q)$ tends to $T_{c}(Q=0)$. Thus despite the nonexistance,
for $Q=0^{+}$,  of a phase transition at or near $T_{c}(Q=0)$, the system is governed, in part, by the proximity to the {\it avoided} critical 
temperature $T_{c}(Q=0)$.

\noindent{\bf Avoided Critical Behaviour To  ${\bf O(n^{-2})}$ :}
We will now examine corrections to the spherical limit.
In a $(1/n)$ expansion$^{[4]}$, the soft term of 
constraint, $[H_{soft}- H_{0}]$ is taken to be small with 
$u = O(1/n) >0$.  The perturbation theory in $u$ is then selectively resummed
treating $1/n$ as the small parameter. As our unperturbed Hamiltonian, we take 
$H_{0}$ in Eq. (3) with a temperature dependent chemical 
potential, $r_{0}+2\sqrt{Q}$, which changes sign at $T=T_{MF}$, 
and is increasingly 
negative at low $T$. 
The Dyson equation implies
\begin{equation}
G^{-1}(\vec{k}) = G^{-1}_{0}(\vec{k}) + \Sigma(\vec{k})
\end{equation}
where, in the continuum limit, $G^{-1}_{0}= r_{0} + k^{2}+ Q k^{-2}$, 
 and $[-\Sigma({\vec  k})]$ is the self-energy.  $T_{c}$, if it exists,
 is determined
 implicitly from the solution of the equation 
 \begin{equation}
 \min_{\vec k} \{ G^{-1}({\vec  k}) \}=0.
 \label{eq:tc}
 \end{equation}
To zeroth order in $1/n$: 
\begin{eqnarray}
G^{-1}({\vec  k})= r+k^{2}+ Q k^{-2}.
\end{eqnarray}
Here $r=r_{0}+\Sigma^{0}$, where $[-\Sigma^{0}]$ denotes the 
self-consistently
computed
$O(n^{0})$ correction to the self-energy. At low temperatures, 
\begin{equation}
\Sigma^{0}(Q,r) \approx \Sigma^{0}(Q=0,r=0)+ \frac{nu}{\pi} 
\sqrt{\frac{Q}{r+2 \sqrt{Q}}},
\end{equation}
where $\Sigma^{0}(Q=0,r=0)$ is the value of the $O(n^{0})$ self-energy
at criticality for the standard three-dimensional ferromagnet
and the second term 
is the $O(n^{0})$ self-energy of a one-dimensional spin chain
with a nearest neighbour exchange interaction proportional to $1/Q$.
Since $\Sigma^{0}$ is manifestly positive, and diverges as
$r\rightarrow r_{\min}=-2 \sqrt{Q}$, Eq. (\ref{eq:tc}) is satisfied 
only in the limit
$r_{0} = -\infty$; to this order
$T_{c}(Q> 0)=0$. We have extended this analysis\cite{[3]} to 
$O(n^{-2})$.  By evaluting diagrams self-consistenly, one observes that all $O(n^{-1})$ and $O(n^{-2})$ self-energy contributions are explicitly
positive, or cancel against more 
divergent positive contributions. We outline here how this is done to 
$O(n^{-1})$. In Fig. 2, the ${\vec k}$-independent $\Sigma^{0}>0$ is the single zeroth order
($O(n^{0})$) contribution. To $O(n^{-1})$ there are two additional diagrams:
$\Sigma^{A}({\vec k})$ and the ${\vec k}$-independent $\Sigma^{B}$.
Inserting, self-consistently, 
$G({\vec  k}) = [G_{0}^{-1}({\vec  k}) + \Sigma^{0}  +\Sigma^{A}({\vec  k})]^{-1}$
into the integral expression $\Sigma^{0} = 
\int \frac{d^{3}k}{(2 \pi)^{3}}~G({\vec  k})$, automatically 
generates $\Sigma^{B}$, as well as higher order diagrams.
This integral diverges as $r \rightarrow r_{\min}$.
The self-energy $\Sigma^{A}$ is positive, and
thus can only further thwart any tendency to order.
A similar anaylsis [5] may be repeated to $O(n^{-2})$.
\begin{figure}[htb]
\centerline{\epsfxsize=3.3in \epsffile{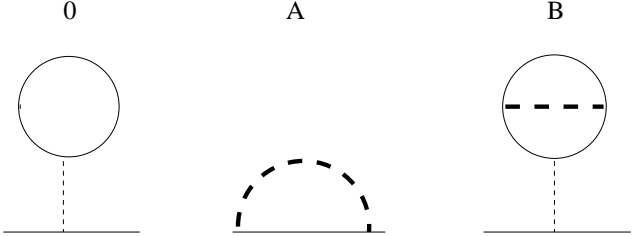}}
\caption{ The Self Energy Corrections. 
The thin dashed lines denote 
bare interaction, the thick dashed lines represent dressed 
interactions
({\it i.e.} bare interactions screened by a geometric series of 
bubble diagrams), and the solid lines denote propagators.} 
\label{fig2}
\end{figure}

All this indicates that to $O(n^{-2})$,  $r=r_{\min}$ is attainable only
at $T_{c}=0$.  Unfortunately, the results of the $1/n$ expansion are
not independent of $Q$ for small $Q$;  we cannot safely draw 
conclussions concerning the $Q\rightarrow 0$ limit as there could, in 
principle, be a change in the behavior of the system when $Q\sim 1/n$.
Nevertheless, the results strongly support the contention that the 
$n \rightarrow \infty$ limit is not singular, and that the spherical 
model captures the important physics of the system for any large $n$.

\noindent{\bf Algebraic Crossover:}  We have also computed the pair 
correlator to $O(n^{-1})$.
At very low temperatures, where $0<r+2 \sqrt{Q} \ll \sqrt{Q}$, 
the 
propagator at intermediate momenta ($1 
\gg |{\vec  k}| \gg Q^{1/4}$) is
\begin{equation}
G^{-1}({\vec  k}) \approx  k^{2}+ [2 \Sigma^{0}/(n^{2}u)] |{\vec  k}| + r 
+ Qk^{-2}. \end{equation} 
We note that
\begin{eqnarray}
G(|{\vec  x}|) 
\sim |{\vec   x}|^{-2} \mbox{ for } \ell_{J} \ll |{\vec   x}| \ll L_{D} 
\ll \xi \nonumber \\
G(|{\vec   x}|) \sim |{\vec  x}|^{-1} \mbox{ for } |{\vec   x}| \ll 
\ell_{J}, 
\end{eqnarray}
where the correlation length
$\ell_{J} \equiv n^{2} u/(2 \Sigma^{0})$, 
is defined in  a way analagous to the Josephson length $^{[6]}$
in the ordered phase of a system with Goldstone modes. 
Thus at sufficiently low temperatures, $T<T_{2}$, 
the (nonoscillatory) spatial behavior of the correlators is again governed by
two length scales. 

{\bf Lattice Effects:}  
The fact that the lattice system has discrete, rather than 
continuous rotational symmetry is reflected in higher order terms
in powers of k, in 
the kernal $J(\vec k)$ in Eq. (\ref{eq:jofk});  the lowest order term
of this sort is $\lambda \sum_{a=1}^{d}k^{4}_{a}$.  The 
effects of these 
terms was determined previously\cite{1} for $n\rightarrow \infty$;  
they produce a
$T_{c}(Q>0)>0$, but the avoided critical phenomena,
{\it i.e.} the fact that $\lim_{Q\rightarrow 0}T_{c}(Q) < 
T_{c}(0)$,
survives for $2< d \le 3$.  More generally, if we apply
the above spin-wave analysis and a Lindemann criterion for $T_{c}$, then 
the same calculation leads to the conclusion that, once again, the 
large $n$ results are qualitatively correct for finite $n$.

\noindent{\bf Multi-spiral states}
Whenever $n \ge 2$, any ground state configuration can be decomposed
into Fourier components,
${\bf S}^{g}({\vec x})= \sum_{i=1}^M \left \{ {\bf a}_i 
\cos[\vec k_{\min}^{(i)}\cdot \vec x] + {\bf b}_i\sin[\vec k_{\min}^{(i)}
\cdot \vec x]
\right \}$ where
$\vec k_{\min}^i$ are chosen from the set of wave vectors
which minimize $J(\vec k)$.
So long as these wave-vectors are ``non-degenerate'', in the sense
that the sum of any pair of wave vectors,  
${\vec k}^{i}_{\min}\pm {\vec k}^{j}_{\min}$  is not equal to
the sum of any other pair of wave vectors, and ``incommensurate'' in 
the sense that for all $i$ and $j$, 
$2({\vec k}^{i}_{\min}+\vec{k}^{j}_{\min})$ is not equal to
a reciprocal lattice vector, it is
straightforward to prove\cite{[3]}
that the condition $[{\bf S}^{g}({\vec x})]^{2}=1$
can be satisfied only if $M \le n/2$.  (These conditions 
are always satisfied for $Q < 4$.)
Thus, for $n\le 3$ 
only simple spiral ($M=1$) ground-states are permitted, while for
$n=4$, a double spiral saturates the bound $^{[12]}$.

\noindent{\bf Acknowledgements:}
We are grateful for many enlightning discussions with 
S.~Chakravarty, V.~J.~Emery, 
D.~Kivelson, and 
G.~Tarjus.  SAK and ZN were supported, in part, 
by NSF grant number DMR-98-08685  at UCLA.  
LNC was supported in part by NSA grant number MDA904-98-1-0518.


%
%

%
%


\begin{references}
\bibitem{[12]} Preliminary calculations
indicate that the generalization of the
expression for $(\Delta S)^{2}$ in Eq. (9) for
$M>1$ is roughly obtained by replacing $(n-2)$ by $(n-2M)$.
\end{references}
\end{document}